\documentclass[10pt,letterpaper,superscriptaddress,email]{article}
\usepackage{opex3}
\usepackage{mathrsfs}
\usepackage{amssymb}
\usepackage{amsmath,amsfonts,latexsym}
\usepackage{array,tabularx,color}
\usepackage[normalem]{ulem}
\usepackage{comment}
\usepackage[amssymb]{SIunits}
\usepackage{bm}
\usepackage{cite}

\begin{document}

\vskip4pc
\title{Role of supercurrents on vortices formation in polariton condensates}

\author{C. Ant\'on,$^{1,*}$ G. Tosi,$^{1}$ M. D. Mart\'in,$^{1}$ L. Vi\~na,$^{1}$ A. Lema\^itre$^{2}$ and J. Bloch$^{2}$}
\address{$^{1}$Departamento de F\'isica de Materiales and Instituto de Ciencia de Materiales "Nicol\'as Cabrera", Universidad
  Aut\'onoma de Madrid, 28049 Madrid, Spain} 
\address{$^{2}$LPN/CNRS, Route de Nozay, 91460, Marcoussis, France}
\email{$^{*}$carlos.anton@uam.es}

\begin{abstract} 
Observation of quantized vortices in non-equilibrium polariton condensates has been reported either by spontaneous formation and pinning in the presence of disorder or by imprinting them onto the signal or idler of an optical parametric oscillator (OPO). Here, we report a detailed analysis of the creation and annihilation of polariton vortex-antivortex pairs in the signal state of a polariton OPO by means of a short optical Gaussian pulse at a certain finite pump wave-vector. A time-resolved, interferometric analysis of the emission allows us to extract the phase of the perturbed condensate and to reveal the dynamics of the supercurrents created by the pulsed probe. This flow is responsible for the appearance of the topological defects when counter-propagating to the underlying currents of the OPO signal.
\end{abstract}



\section{Introduction}

	The physics of strong radiation-matter interaction in semiconductor microcavities has been intensively studied in the last two decades ~\cite{Weisbuch92,kavokin07}. Under these strong coupling conditions, new quasi-particles, exciton-polaritons are formed as a result of the quantum superposition of exciton and photon states. These polaritons possess a very peculiar dispersion relation, which allows parametric instabilities and amplification that a quadratic dispersion would not allow. Furthermore, they maintain not only the low mass and reduced density of states of their photonic part but also the capacity to interact with their surroundings from their excitonic content. These properties make them very suitable to obtain a Bose-Einstein condensation similar to that observed in dilute atomic gases ~\cite{anderson95}.

	Only recently, polariton condensation was experimentally confirmed ~\cite{kasprzak06} and it paved the way to the study of many collective quantum phenomena in semiconductor microcavities, such as superfluid properties ~\cite{amo09,amo09a,amo11,Nardin11,Grosso11,Sich12}, or macroscopic phase coherence ~\cite{Krizhanovskii09,spano11,carusotto05,Wouters06,Manni11}. Many of these results were obtained after quasi-resonant excitation at the inflection point of the lower polariton branch (LPB), in the so-called optical parametric oscillator (OPO) configuration. Under these pumping conditions, polaritons undergo stimulated parametric scattering, giving rise to the build-up of signal and idler polariton populations, which eventually condense. 
	
	Of special interest for our work here is the behaviour of polariton droplets, created optically by an extra pulsed-probe beam, at the signal state of an OPO. The pioneering work of Amo and coworkers ~\cite{amo09} described the superfluid-like motion across a condensate of such drops. 
	
	Among the collective phenomena in polariton condensates, the appearance of quantized vortices deserves special attention since they are regarded as a fingerprint of superfluid behavior ~\cite{onsager_49,feynman_55}; they are topological defects that occur on macroscopically coherent systems. The phase of the condensate winds from $0$ to $2 m \pi$ around their cores, evidencing that the vortices carry a quantized angular momentum $m\hbar$. These vortices have been observed in semiconductor microcavities only recently~\cite{krizhanovskii10,sanvitto10,tosi10,roumpos10}. However, the correspondence between vortices and superfluid behaviour is not always valid, as observed in polariton condensates where they can also appear as a consequence of natural defects in the samples and the competition between gain and decay processes ~\cite{lagoudakis11,lagoudakis08,lagoudakis09a,marchetti12,keeling08,borgh10}. 
	
	The unintended formation of a $m=-1/+1$, antivortex/vortex,  when imprinting a $m=+1/-1$, vortex/antivortex, in the OPO-signal has been reported recently~\cite{tosi11}, where it was also shown how to induce vortex-antivortex (V-AV) pairs just with a drop formed with a Gaussian-probe. In this work, we analyze the dynamics of these pairs created by a Gaussian beam in terms of the experimentally-deduced supercurrents in the condensate: we experimentally demonstrate that a finite difference between the OPO-signal and probe momenta, $\mathbf{k}_{s}-\mathbf{k}_{pb}$, is essential for the V-AV pair formation. Furthermore, we also evidence the existence of dissipation and that these vortices do not imply superfluidity.

\section{Sample and experimental setup}

\begin{figure}[!htb]
\begin{center}
\includegraphics[width=7.0 cm]{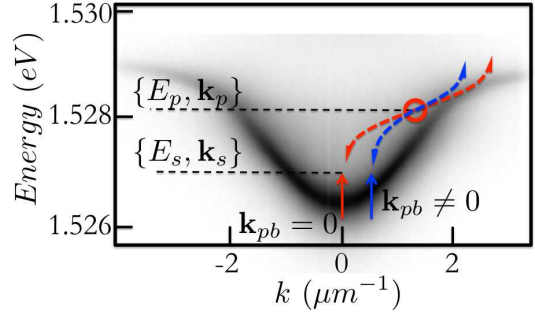}
\vspace{-0.5 cm}
\end{center}
\caption{(Color online) Experimental lower-polariton-branch dispersion relation under non-resonant weak excitation conditions, energy versus in-plane $\mathbf{k}$ momentum. The blue-shifted conditions used for the OPO-pump and pulsed-probe are: pump laser $E_p = 1.5283$~eV and $k_{p}=1.4$~$\mu$m$^{-1}$ (red ring); probe beam $E_s= 1.5268$~eV either with $k_{pb}=0$ or $k_{pb}=0.5$~$\mu$m$^{-1}$ (red/blue dashed arrows depict the corresponding pair-scattering process).}
\label{fig:fig1}
\end{figure}

	We study a $\lambda / 2$ AlAs microcavity with a single GaAs quantum well placed at the antinode of the mirror-confined cavity field, characterized by a Rabi splitting of $4.4$~meV ~\cite{perrin05}. Figure~\ref{fig:fig1} sketches our experimental excitation conditions. Maintaining the sample at $10$~K, a cw $Ti:Al_2 O_3$ laser resonantly pumps polaritons at $E_p=1.5283$~eV and  $k_{p}=1.4$~$\mu$m$^{-1}$, close to the inflection point of the LPB dispersion (a small blue-shift allows fulfilling of the OPO processes). Above the threshold for parametric scattering ($250$~W$/$cm$^2$), the system enters the steady state OPO regime. The laser power is set slightly above threshold for the generation of condensed polariton states at $E_s= 1.5268$~eV and $\mathbf{k}_s \simeq \mathbf{0}$ (OPO-signal) and $E_i = 2E_p - E_s$ and $\mathbf{k}_i \simeq 2\mathbf{k}_p$ (OPO-idler). The OPO-signal is $\simeq 1$~meV blue-shifted from the bottom of the LP bare dispersion. A negative detuning ~\cite{detuning},  $\simeq -1$~meV, is chosen for the experiments. The condensate extends over an area of $\sim 50 \times 50$~$\mu$m$^2$. To study the dynamics of topological defects creation and its relation with the supercurrents, we shine a $2$~ps-long Gaussian pulsed beam in resonance with the OPO-signal at $E_s$ with an area of $\sim 90$~$\mu$m$^2$ at two different angles, i.e. different momenta, $k_{pb}=0$ and $k_{pb}=0.5$~$\mu$m$^{-1}$, and a typical power of $3$~$\mu$W. 
	
	Intensity maps of the emission are obtained over the condensate area. The intensity is averaged over millions of shots with a streak camera, implying that our results correspond to deterministic dynamics of the system (random events statistically disappear). In order to highlight the modifications in the polariton spatial density introduced by the probe, we present images were the background steady state emission is subtracted, and we dub this information as gain. To avoid the direct pump laser light, we filter the emission in $\mathbf{k}$-space around $\mathbf{k}_{s}$ and we follow the evolution of the gain after the arrival of the Gaussian-probe. To analyze the dynamics of the phase in the polariton condensates, interference images between the signal emission and an expanded, constant phase, region of it are obtained in a Mach-Zehnder interferometer. 


\section{Data analysis}
	
\begin{figure}[!htb]
\begin{center}
\includegraphics[width=11.0 cm]{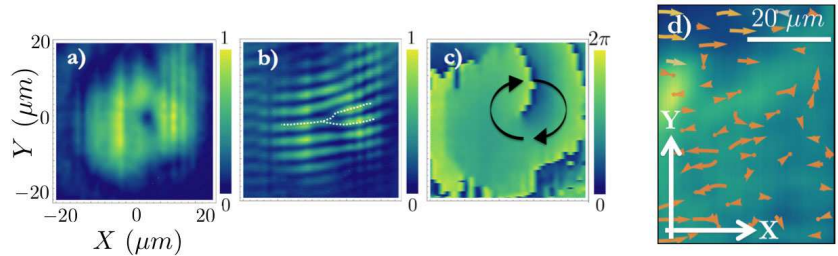}
\vspace{-0.5 cm}
\end{center}
\caption{(Color online) (a) Signal emission $\left|\mathscr{E}_{s}\right|^2$ recorded from the unexpanded interferometer arm, showing a typical vortex. (b) Interference pattern $\left|\mathscr{E}_s+\mathscr{E}_r\right|^2$ with a fork-like dislocation that reveals the position of the vortex (white dotted line is a guide to the eye). (c) Extracted phase, $\Phi_s$, with the Fourier procedure detailed in the text, black arrows schematically depict polariton circulation around the vortex core. The same sample position and scale are used for the three images. 
(d) Supercurrents stream-plot (orange arrows) of an unperturbed OPO-signal superimposed on its emission shown in a false color scale.}
\label{fig:vortex_example}
\end{figure}

	An interferometric analysis unambiguously reveals the presence of topological defects and obtains the supercurrents in the condensate. A polariton vortex shows a doughnut-like emission as that depicted in Fig. ~\ref{fig:vortex_example} (a), but it is necessary to interfere its emission with a coherent wave to confirm that the hole at its center is due to a vortex core. The appearance of fork-like dislocations on the interference pattern with a difference of $m$ fringes corresponds to a phase winding by $2 \pi m$ around the vortex core,  Fig. ~\ref{fig:vortex_example} (b), ~\cite{lagoudakis11,lagoudakis08,lagoudakis09a,tosi11}. We perform a Fourier analysis of the interferometer images to obtain the phase maps of the condensate, Fig. ~\ref{fig:vortex_example} (c). The interference of the OPO-signal emission, denoted as $\mathscr{E}_s\equiv |\mathscr{E}_{s}|e^{i\Phi_s}$, with an expanded part of itself, $\mathscr{E}_r\equiv |\mathscr{E}_{r}|e^{i\Phi_r}$, that can be considered as a plane wave, is registered in the Mach-Zehnder interferometer. 
	
	The detected interference can be written as:
\begin{equation}
\left|\mathscr{E}_s+\mathscr{E}_r\right|^2=\left(\left| \mathscr{E}_s\right|^2+\left| \mathscr{E}_r\right|^2\right)+ \mathscr{E}_s^\ast \mathscr{E}_r+ \mathscr{E}_r^\ast \mathscr{E}_s\; .
\label{eq:eqintens} 
\end{equation}
	Performing a Fourier transform from real ($\mathbf{r}$) to reciprocal ($\mathbf{k}$) space, $\mathscr{F} (\left|\mathscr{E}_s+\mathscr{E}_r\right|^2)$, three contributions are obtained ~\cite{Liebling2004}: the term $\mathscr{F} ( \left| \mathscr{E}_s\right|^2+\left| \mathscr{E}_r\right|^2 )$ is centered at the Fourier space origin; $\mathscr{E}_s$ and $\mathscr{E}_r$ interfere at the detector with a finite angle, the interference mixing terms, $\mathscr{F} ( \mathscr{E}_s^\ast \mathscr{E}_r )$ and $\mathscr{F} ( \mathscr{E}_r^\ast \mathscr{E}_s)$, appear in opposite symmetrical places as off-axis contributions. We shift the whole Fourier map, placing $\mathscr{F} ( \mathscr{E}_r^\ast \mathscr{E}_s )$ at the origin, and we apply a top-hat filter to only keep this term. 
	
	Finally, by performing an inverse Fourier transform, we recover a complex matrix that is proportional to the amplitude and phase of the emission and therefore to the polariton wave function, $\mathscr{E}_s=\left| \mathscr{E}_{s} \right| e^{i \Phi_s}$, providing full information about the phase of the condensate $\Phi_s \left( \mathbf{r}, t \right)$ (Fig. ~\ref{fig:vortex_example} c). From the phase field it is also possible to extract experimentally the polariton supercurrents. For a complex field $\left|\mathscr{E} \left(\mathbf{r}, t \right)\right| e^{i \Phi \left( \mathbf{r}, t \right)}$, describing a macroscopic number of particles condensed in the same quantum state, the current is defined as, ~\cite{marchetti12}
\begin{equation}
\mathbf{j} \left( \mathbf{r} , t \right) = \frac{\hbar}{m} \left| \mathscr{E} \left(\mathbf{r}, t \right)\right|^2 \nabla  \Phi \left( \mathbf{r}, t \right) =  \left| \mathscr{E} \left(\mathbf{r}, t \right)\right|^2 \mathbf{v}_s \left( \mathbf{r}, t \right) \; ,
\label{ec:current} 
\end{equation}
\noindent where $\mathbf{v}_s \left( \mathbf{r}, t \right)$ represents the flow velocity. Here, we will refer to the supercurrents as the gradient of the phase only, $\nabla  \Phi \left( \mathbf{r}, t \right)$. 
		
	Figure ~\ref{fig:vortex_example} (d) presents, by means of orange arrows, such a supercurrents stream-plot for the case of an unperturbed OPO condensate, obtained as described in Sect. 2, with its emission shown in a false color scale on the background.~\cite{scnote} The supercurrents do not show a privileged direction on this situation.
		

\section{Results}

	We analyze two different Gaussian-probe configurations, $k_{pb} = 0.5$~$\mu$m$^{-1}$ and $k_{pb} = 0$, which develop different effects on the OPO-polariton signal dynamics.
	
\begin{figure}[!hbt]
\begin{center}
\includegraphics[width=9.0 cm]{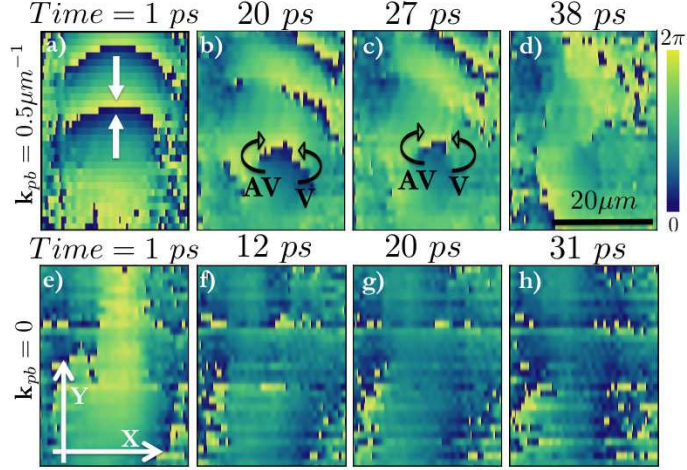}
\vspace{-0.5 cm}
\end{center}
\caption{(Color online) (a-d)/(e-h) Phase maps at different times after the probe disappearance for $k_{pb} = 0.5$/$k_{pb} = 0$~$\mu$m$^{-1}$. Same false color scale for all plots from $0$ to $2 \pi$. The white arrows in (a) point to a phase jump between $0$ and $2\pi$. The black curved arrows in (b,c) depict the circulation around the V and AV cores.}
\label{fig:phasevst}
\end{figure}

	We start analyzing the case of $k_{pb} = 0.5$~$\mu$m$^{-1}$ (blue arrows in Fig. ~\ref{fig:fig1}). The phase maps at different times after the excitation are shown in Fig. ~\ref{fig:phasevst} (a-d). One picosecond after the probe is gone, Fig. ~\ref{fig:phasevst} (a), there is a slope in phase observed as a sawtooth function, bounded by $mod \left(2 \pi \right)$. However, the phase must be considered as a continuously increasing function in this snapshot, avoiding the $0$-$2\pi$ discontinuities ~\cite{marchetti12}. The pulsed probe creates a moving ellipsoidal polariton bullet, as will be shown below, and leads to the creation of V-AV pairs at its borders. One of this pairs, which we study in detail, is seen in Fig. ~\ref{fig:phasevst} (b) at $\sim 20$~ps and it is marked with black arrows. The circulation is anticlockwise (clockwise) for the vortex (antivortex) on the right (left) side of the map, Fig. ~\ref{fig:phasevst}(b,c). These snapshots at $t=20$~ps and $t=27$~ps reveal a progressive approach of the V-AV. The pair self-annihilates after $\sim 40$~ps, as clearly seen in Fig. ~\ref{fig:phasevst}(d), which becomes free from topological defects. 

	Now, we consider the instance of $k_{pb} = 0$ (red arrows in Fig. ~\ref{fig:fig1}). In order to compare properly the differences in the phase dynamics, the same region of interest has been chosen for the snapshots compiled in Figs. \ref{fig:phasevst} (a-d) and (e-h). The probe at $k_{pb} = 0$ does not change significantly the OPO-signal phase as demonstrated in Figs. ~\ref{fig:phasevst} (e-h): no V-AV pairs appear in this case. However, the phase map is determined by the potential landscape, i. e. local defects produce different phase values, and therefore it is considerably dependent on the details of the sample morphology in the region used for the experiments.
	
\begin{figure*}[!htb]
\begin{center}
\includegraphics[width=12.0 cm]{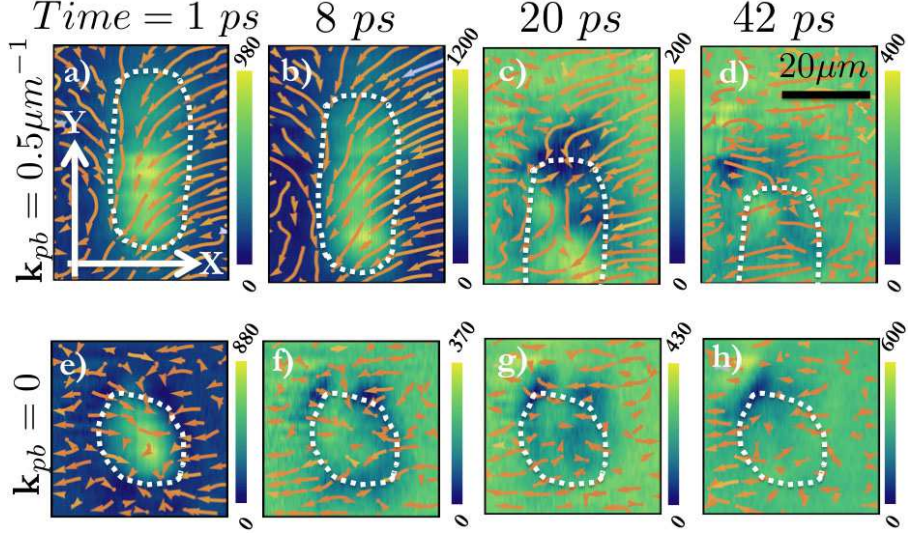}
\vspace{-0.5 cm}
\end{center}
\caption{(Color online) Supercurrents stream-plot superimposed on the gain at the OPO-signal (orange arrows). (a-d) Evolution of the polariton bullet at $k_{pb} = 0.5$~$\mu$m$^{-1}$. The dotted white line is a guide to the eye to follow the movement of the bullet. Spatial extension of each snapshot is $40 \times 55$~$\mu$ m$^2$. (e-h) Same as (a-d) for $k_{pb} = 0$, the probe size is sketched by the dotted white line. Spatial extension of each snapshot is $40 \times 40$~$\mu$m$^2$. The gain in each figure is rescaled according to its false color bar.}

\label{fig:scemvst}
\end{figure*}

	Besides the phase information, our experiments also yield the gain density distribution in real space directly from the emission intensities. Its time evolution is displayed in a false color scale in Figs. ~\ref{fig:scemvst} (a-d) for $k_{pb} = 0.5$~$\mu$m$^{-1}$ together with the supercurrents (orange arrows), extracted from the phase as explained above. This gain (delimited in the figure with white dotted lines) constitutes a moving polariton droplet, which propagates $\sim 30 $~$\mu$m in the negative $Y$-direction, as imposed by the geometry of the probe impinging on the sample surface. Its intensity decays, as the bullet travels downwards, due to emission processes and to diffusion of some polaritons towards the full area of the OPO-signal (note that the area outside the white lines becomes brighter). The bullet propagation involves a net supercurrent flow in the negative $Y$-direction as demonstrated by the stream-plot on the top of the emission images. After $\sim 30$~ps the bullet dissolves due to friction and polariton diffusion. The supercurrents reach a state in which probe effects have completely disappeared. 
	
	From our previous work, where vortices were imprinted with Gauss-Laguerrian beams and unintended antivortices appeared at given locations at the borders of the perturbing beam ~\cite{tosi11}, we know that these locations are determined by counter-propagating currents of the underlying OPO-signal and those created by the probe beams. Therefore, in our present measurements, we can infer that in the positions where the V-AV pairs appear  (see Fig. ~\ref{fig:phasevst} (b)) there must be underlying currents always counter-propagating to those currents shown in Fig. ~\ref{fig:scemvst} (b). This fact leads to the deterministic appearance of the topological defects in experiments that average over millions of shots in the streak camera. It must be mentioned that in Figs. ~\ref{fig:scemvst} (b-c) the presence of V-AV pairs should be apparent as close circulations of the supercurrents, however they are not observed due to the spatial resolution of our setup.
	
	Figures ~\ref{fig:scemvst} (e-h) show the gain for $k_{pb} = 0$. Maps of supercurrents without a preferred direction over a stopped polariton droplet, which rapidly dissolves in the OPO-signal, are observed. In this situation, topological defects are not created. One could try to find whether there is a critical value of the probe momentum for the formation of V-AV pairs, $\mathbf{k}_{pb}^{crit.}$, but this is not an easy task since $\mathbf{k}_{pb}^{crit.}$ is determined by many parameters, some of them controllable as power densities of pump and probe beams, size of the beams, etc, but others not manageable as, for example, sample morphology and natural defects.

\begin{figure}[!htb]
\begin{center}
\includegraphics[width=7.0 cm]{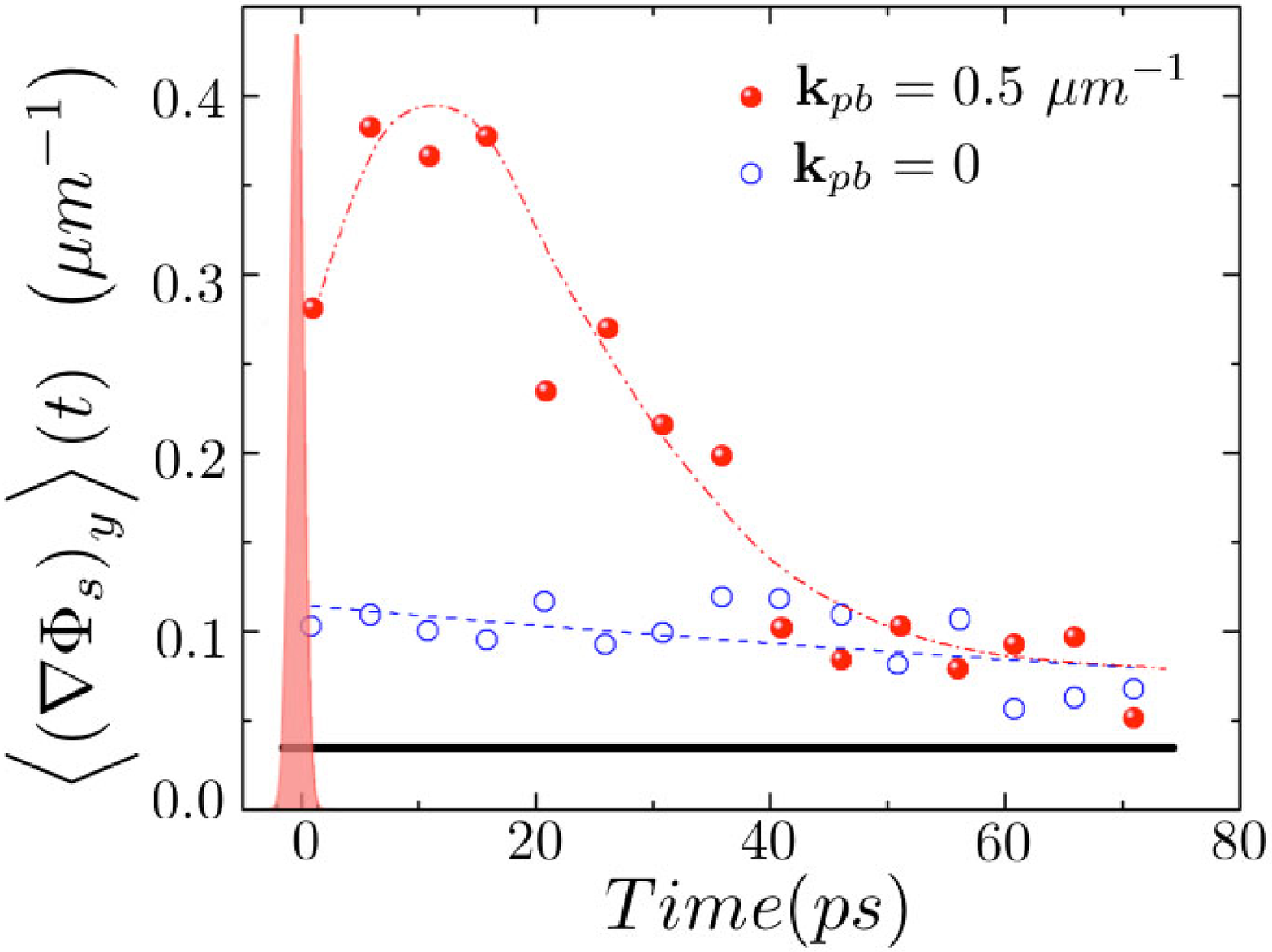}
\vspace{-0.5 cm}
\end{center}
\caption{(Color online) Time evolution of the mean supercurrents along the $Y$-direction under two probe configurations. Solid dots (open dots) correspond to $k_{pb} = 0.5$~$\mu$m$^{-1}$ ($k_{pb} = 0$) with a dot-dashed (dashed) line as a guide to the eyes. The red gaussian peak depicts the $2$ ps-pulsed probe arrival. The full black line indicates the value of the mean current along the $Y$-axis in the steady state OPO-signal (as obtained from Fig. ~\ref{fig:vortex_example} (d)).}
\label{fig:scvst}
\end{figure}

	With the choice of coordinates shown in Fig. ~\ref{fig:scemvst}(a), the arrival of the probe with $\mathbf{k}_{pb}=-\left(0.5\right) \mathbf{u}_y$~$\mu$m$^{-1}$ makes the $Y$-direction a privileged one in which the polariton population moves. To obtain a deeper insight on the supercurrents dynamics, we quantify the time-evolution of their mean value along this direction in the images shown in Fig. ~\ref{fig:scemvst}. We assign a supercurrent vector, $\mathbf{\nabla} \Phi_s  \left(\mathbf{r},t\right) = \left(\mathbf{\nabla} \Phi_s \left(\mathbf{r},t\right) \right)_x  \mathbf{u}_x+\left(\mathbf{\nabla} \Phi_s \left(\mathbf{r},t\right) \right)_y \mathbf{u}_y$, to each pixel ($l=1,...,N_x$ and $m =1,...,N_y$) of the snapshots and we compute the average of its $Y$-component: 
\begin{equation}
\left\langle \left(\mathbf{\nabla} \Phi_s \right)_y \right\rangle \left(t\right) \equiv \displaystyle\frac{1}{N_x N_y} \displaystyle\sum_{l,m=1}^{N_x,N_y} \left(\mathbf{\nabla} \Phi_s \left(\left(l,m\right),t\right) \right)_y\;,
\label{eq:eqsc} 
\end{equation}	
which measures the evolution of the mean currents along the $Y$-direction. The dynamics for the two cases studied in this work are very different as demonstrated in Fig.  ~\ref{fig:scvst}, which depicts $|\langle \left(\mathbf{\nabla} \Phi_s \right)_y \rangle \left(t\right)|$ for $k_{pb} = 0.5$~$\mu$m$^{-1}$ (solid dots) and $k_{pb} = 0$ (open dots). In the former case, an initial increase lasting $\sim 10$~ps is followed by a decrease with a characteristic decay time of $\sim 20$~ps. The maximum value is close to that of the probe beam, $\mathbf{k}_{pb}$. The decay evidences the slow-down of the polariton bullet due to friction and its consequently solution on the steady-state OPO-signal population. This demonstrates that, for our experimental conditions and in the chosen sample region, a superfluid-like behaviour as that previously reported ~\cite{amo09,amo09a} is not achieved. For $k_{pb} = 0$, $|\langle \left(\mathbf{\nabla} \Phi_s \right)_y \rangle \left(t\right)|$ decays slowly from an initial value of $0.11$~$\mu$m$^{-1}$. The values of $|\langle \left(\mathbf{\nabla} \Phi_s \right)_y \rangle \left(t\right)|$ at long times for $k_{pb} \neq 0$ and $k_{pb} = 0$ coincide and they represent the diffusion out of the droplet, in the $-Y$-direction, of the extra population created by the probe.
	
	It is important to remark that the mean value, $|\langle \left(\mathbf{\nabla} \Phi_s \right)_y \rangle|$, for the cw OPO-signal shown in Fig. ~\ref{fig:vortex_example} (d) is $0.03$~$\mu$m$^{-1}$ (full line in Fig. ~\ref{fig:scvst}), considerably smaller than those shown for the perturbed cases. The fact that the difference between the underlying cw-currents and those generated by the $k_{pb} = 0$ probe is not enough to create V-AV pairs indicates that a threshold superior to $(0.11-0.03)$~$\mu$m$^{-1}$ $\approx 0.08$~$\mu$m$^{-1}$ is required to observe vortices for the method employed in our experiments. 
	
\section{Conclusions} 

	Vortices in OPO-polariton condensates can be generated as a response to counter-propagation of supercurrents in the cw OPO-signal and in the extra population induced by a Gaussian-probe beam at a certain finite $\mathbf{k}_{pb}$. A full analysis of the phase of the condensates, perturbed by the probe, yields the dynamics of formation and annihilation of V-AV pairs and of the supercurrents. For a steady state OPO-signal the polariton supercurrents are randomly distributed. A certain threshold for the difference between the supercurrents is needed for the observation of topological defects.
	
\section{Acknowledgements}
	
	We thank F. M. Marchetti, F.P. Laussy and C. Tejedor for fruitful discussion. G. T.  and C.A. acknowledge financial support from FPI and FPU scholarship, respectively. Work was supported by the Spanish MEC MAT2011-22997, CAM (S-2009/ESP-1503) and FP7 ITN Clermont4 (235114).


\begin{thebibliography}{99}
 
\bibitem{Weisbuch92} C. Weisbuch, M. Nishioka, A. Ishikawa, and Y. Arakawa, "Observation of the coupled exciton-photon mode splitting in a semiconductor quantum microcavity" \prl {\bf 69}, 3314-3317 (1992).

\bibitem{kavokin07} A. Kavokin, J. J. Baumberg, G. Malpuech and F.P. Laussy, "Microcavities", Oxford University Press (2007).

\bibitem{anderson95} M. H. Anderson, J. R. Ensher, M. R. Matthews, C. E. Wieman, E. A. Cornell, "Observation of Bose-Einstein Condensation in a Dilute Atomic Vapor", Science \textbf{ 269}, 5221, 198-201 (1995).

\bibitem{kasprzak06} J. Kasprzak, M. Richard, S. Kundermann, A. Baas, P. Jeambrun, J. M. J. Keeling, F. M. Marchetti, M. H. Szyma\'nska, R. AndrŽ, J. L. Staehli, V. Savona, P. B. Littlewood, B. Deveaud and Le Si Dang, "Bose-Einstein condensation of exciton polaritons", \nat {\bf 443}, 409 - 412 (2006).

\bibitem{amo09} A. Amo, D. Sanvitto, F. P. Laussy, D. Ballarini, E. del Valle, M. D. Martin, A. Lem\^itre, J. Bloch, D. N. Krizhanovskii, M. S. Skolnick, C. Tejedor and L. Vi\~na, "Collective fluid dynamics of a polariton condensate in a semiconductor microcavity", \nat {\bf 457}, 291 - 295 (2009).

\bibitem{amo09a} A. Amo, J. Lefr\`ere, S. Pigeon, C. Adrados, C. Ciuti, I. Carusotto, R. Houdr\'e, E. Giacobino and A. Bramati, "Superfluidity of polaritons in semiconductor microcavities", Nature Phys. \textbf{5}, 805 - 810 (2009).

\bibitem{amo11} A. Amo, S. Pigeon, D. Sanvitto, V. G. Sala, R. Hivet, I. Carusotto, F. Pisanello, G. Lem\'enager, R. Houdr\'e, E Giacobino, C. Ciuti and A. Bramati, "Polariton Superfluids Reveal Quantum Hydrodynamic Solitons", Science, \textbf{332}, 1167-1170 (2011).

\bibitem{Grosso11} G. Grosso, G. Nardin, F. Morier-Genoud, Y. L\'eger, and B. Deveaud-Pl\'edran, "Soliton Instabilities and Vortex Street Formation in a Polariton Quantum Fluid", \prl {\bf 107}, 245301 (2011).

\bibitem{Nardin11} G. Nardin, G. Grosso, Y. L\'eger, B. Pi\c{e}tka, F. Morier-Genoud and B. Deveaud-Pl\'edran, "Hydrodynamic nucleation of quantized vortex pairs in a polariton quantum fluid", Nat. Phys. \textbf{7}, 635 - 641 (2011).

\bibitem{Sich12} M. Sich, D. N. Krizhanovskii, M. S. Skolnick, A. V. Gorbach, R. Hartley, D. V. Skryabin, E. A. Cerda-M\'endez, K. Biermann, R. Hey and P. V. Santos, "Observation of bright polariton solitons in a semiconductor microcavity", Nat. Phot. \textbf{6}, 50-55 (2012).

\bibitem{Krizhanovskii09} D. N. Krizhanovskii, K. G. Lagoudakis, M. Wouters, B. Pietka, R. A. Bradley, K. Guda, D. M. Whittaker, M. S. Skolnick, B. Deveaud-Pl\'edran, M. Richard, R. Andr\'e, and Le Si Dang , "Coexisting nonequilibrium condensates with long-range spatial coherence in semiconductor microcavities", \prb {\bf  80}, 045317 (2009).  

\bibitem{spano11} R. Spano, J. Cuadra, C. Lingg, D. Sanvitto, M. D. Mart\'in, P. R. Eastham, M. van der Poel, J. M. Hvam, L. Vi\~na, "Build up of off-diagonal long-range order in microcavity exciton-polaritons across the parametric threshold", arXiv:1111.4894 (2011).

 \bibitem{carusotto05} I. Carusotto and C. Ciuti, "Spontaneous microcavity-polariton coherence across the parametric threshold: Quantum Monte Carlo studies", \prb {\bf 72}, 125335 (2005).

\bibitem{Wouters06} M. Wouters, I. Carusotto, "Absence of long-range coherence in the parametric emission of photonic wires", \prb {\bf 74}, 245316 (2006).

\bibitem{Manni11} F. Manni, K. G. Lagoudakis, B. Pi\c{e}tka, L. Fontanesi, M. Wouters, V. Savona, R. Andr\'e, B. Deveaud-Pl\'edran, "Polariton Condensation in a One-Dimensional Disordered Potential", \prl {\bf 106}, 176401 (2011).

\bibitem{onsager_49} L. Onsager, "Statistical Hidrodynamics", Nuovo Cimento \textbf{6}, Suppl. 2, 249 (1949).

\bibitem{feynman_55} R. Feynman, "Vol. 1 of Progress in Low Temperature Physics", Elsevier, 17-53 (1955).

\bibitem{krizhanovskii10} D. N. Krizhanovskii, D. M. Whittaker, R. A. Bradley, K. Guda, D. Sarkar, D. Sanvitto, L. Vi\~na, E. Cerda, P. Santos, K. Biermann, R. Hey, and M. S. Skolnick, "Effect of Interactions on Vortices in a Nonequilibrium Polariton Condensate", \prl {\bf 104}, 126402 (2010).

\bibitem{sanvitto10} D. Sanvitto, F. M. Marchetti, M. H. Szyma\'nska, G. Tosi, M. Baudisch, F. P. Laussy, D. N. Krizhanovskii, M. S. Skolnick, L. Marrucci, A. Lema"tre, J. Bloch, C. Tejedor and L. Vi\~na, "Persistent currents and quantized vortices in a polariton superfluid", Nature Phys. \textbf{6}, 527 - 533 (2010).

\bibitem{tosi10} G. Tosi, M. Baudisch, D. Sanvitto, L. Vi\~na, A Lema\^itre, J. Bloch, E. Karimi, B. Piccirillo and L. Marrucci, "Optical induced vortices and persistent currents in polariton condensates", J. Phys.: Conf. Ser. \textbf{210}, 012023 (2010).

\bibitem{roumpos10} G. Roumpos, M. D. Fraser, A. L\"offler, S. H\"ofling, A. Forchel and Y. Yamamoto, "Single vortex-antivortex pair in an exciton-polariton condensate", Nature Phys.  \textbf{7}, 129 - 133 (2010).

\bibitem{lagoudakis11} K. G. Lagoudakis, F. Manni, B. Pi\c{e}tka, M. Wouters, T. C. H. Liew, V. Savona, A. V. Kavokin, R. Andr\'e, and B. Deveaud-Pl\'edran, "Probing the Dynamics of Spontaneous Quantum Vortices in Polariton Superfluids", \prl {\bf 106}, 115301 (2011).

\bibitem{lagoudakis08} K. G. Lagoudakis, M. Wouters, M. Richard, A. Baas, I. Carusotto, R. Andr\'e, L. S. Dang and B. Deveaud-Pl\'edran, "Quantized vortices in an exciton-polariton condensate", Nature Phys. \textbf{4}, 706 - 710 (2008).

\bibitem{lagoudakis09a} K. G. Lagoudakis, T. Ostatnick\'y, A. V. Kavokin, Y. G. Rubo, R. Andr\'e and B. Deveaud-Pl\'edran, "Observation of Half-Quantum Vortices in an Exciton-Polariton Condensate", Science \textbf{326}, 974 - 976  (2009).

\bibitem{marchetti12} F. M. Marchetti, M. H. Szyma\'nska, "Vortices in polariton OPO superfluids", arXiv:1107.4487 (2012).

\bibitem{keeling08} J. Keeling and N. G. Berloff, "Spontaneous Rotating Vortex Lattices in a Pumped Decaying Condensate", \prl {\bf 100}, 250401 (2008).

\bibitem{borgh10} M. O. Borgh, J. Keeling, and N. G. Berloff, "Spatial pattern formation and polarization dynamics of a nonequilibrium spinor polariton condensate", \prb {\bf 81}, 235302 (2010).

\bibitem{tosi11} G. Tosi, F. M. Marchetti, D. Sanvitto, C. Ant\'on, M. H. Szyma\'nska, A. Berceanu, C. Tejedor, L. Marrucci, A. Lema\^itre, J. Bloch, and L. Vi\~na, "Onset and Dynamics of Vortex-Antivortex Pairs in Polariton Optical Parametric Oscillator Superfluids", \prl {\bf 107}, 036401 (2011).

\bibitem{perrin05} M. Perrin, P. Senellart, A. Lema\^itre, and J. Bloch, "Polariton relaxation in semiconductor microcavities: Efficiency of electron-polariton scattering", \prb {\bf 72}, 075340 (2005).

\bibitem{detuning} The detuning $\delta$ is defined as the difference between the bare cavity mode, $E_C$, and exciton, $E_X$, energies at $\mathbf{k} = \mathbf{0}$.

\bibitem{Liebling2004} M. Liebling, T. Blu and M. Unser, "Complex-wave retrieval from a single off-axis hologram", \josaa {\bf 21}, 367 (2004).

\bibitem{scnote} For the supercurrent calculations, eventual phase jumps between $0$ and $2\pi$, where the gradient of the phase is not well defined, (see, for example the region between the white arrows in the Fig. ~\ref{fig:phasevst} (a)), are circumvented using $\nabla \Phi_s \left( \mathbf{r}, t \right) = -i \nabla \exp\left({i \Phi_s \left( \mathbf{r}, t \right)}\right)/ \exp \left({i \Phi_s \left( \mathbf{r}, t \right)}\right)$.

\end{thebibliography}
\end{document}